\documentclass[a4]{elsart5p}

\usepackage{amssymb,amsmath} 
\usepackage{bm,hyphenat,xspace} 
\usepackage{graphicx,epsfig}
\usepackage{tabularx}
\usepackage{epsfig,pstricks,psfig,epic,eepic,pst-coil}

\newcommand{\email}[1]{\ead{#1}}

\newcommand{\He}{{}^3\mathrm{He}}

\begin{document}

\begin{frontmatter}

\title {Signatures of the $\Delta$ isobar in spin observables of ${}^3\mathrm{He}$ electrodisintegration}

\author{A.~Deltuva}
\email{arnoldas.deltuva@tfai.vu.lt}

\address{
Institute of Theoretical Physics and Astronomy, 
Vilnius University, Saul\.etekio al. 3, LT-10257 Vilnius, Lithuania}



\begin{abstract}
  The electrodisintegration of $\He$ is considered focusing  on the effects
  of the $\Delta$ isobar excitation which is treated dynamically on the same footing as nucleons.
  In the region beyond the quasi-elastic peak
the  predicted transverse response functions   $R_{T}$ and  $R_{T'}$ are visibly affected. This
leads to sizable  $\Delta$ isobar effects for inclusive and exclusive
electron polarization asymmetries in particular kinematic regions.
A measurement performed in the proposed regime could provide judgment
for  models of nuclear forces and currents.
\end{abstract}


\end{frontmatter}


\section{Introduction \label{sec:intro}}

The $\Delta$ isobar plays an important role in the nuclear dynamics.
It manifests itself most prominently as the $P$-wave pion-nucleon resonance in the
spin-isospin $\frac32$ channel around the total energy of 1232 MeV. In the intermediate
energy regime the $\Delta$ isobar is a dominant mechanism for the pion production and
absorption in hadronic and electromagnetic processes \cite{PASCALUTSA2007125}.
However, even at considerably lower energies virtual excitations of nucleons ($N$) to
$\Delta$ isobars yield important contributions to nuclear forces and electroweak
currents. One of the most famous examples is the Fujita-Miyazawa three-nucleon force
\cite{fujita:57a}. The majority of works describing the processes in systems with three or more
nucleons relied on the static approximation for the $\Delta$ isobar, that is,
restricting the Hilbert space to  purely nucleonic degrees of freedom while 
accounting for leading $\Delta$ isobar contributions via effective nucleonic
forces and current operators 
\cite{carlson:98a,marcucci:05a,YUAN201190}. 
This simplifies the description of the few-nucleon problem but is not always reliable
with respect to the $\Delta$ isobar effect \cite{sauer:86a}.
In contrast, the Hannover group pursued an alternative description where the 
$\Delta$ isobar was an active degree of freedom  treated on the same footing as nucleons
\cite{sauer:86a,hajduk:79a,hajduk:83a,nemoto:98b}.
This leads to a more complicated Hilbert space with the channel coupling,
but avoids the static approximation and offers the possibility to include more rich dynamics
in terms of $\Delta$ isobar contributions, such as those mediated by heavier mesons.

Reactions in the three-nucleon system  at energies below or around the pion-production
threshold have been described considering the $\Delta$ isobar as a stable baryon of
spin and isospin $\frac32$, since in this energy regime only a virtual
excitation of a nucleon to a $\Delta$ isobar is possible.
Prominent $\Delta$ isobar effects have been found in the differential cross section
and some spin observables of the nucleon-deuteron elastic scattering and breakup
and radiative capture \cite{deltuva:03c,deltuva:04a}. Compared to those reactions,
the electrodisintegration of the trinucleon bound state offers in addition the opportunity
to test the models for the electromagnetic current at higher four-momentum transfers.
However, the study of \cite{deltuva:04b} found only small  $\Delta$ isobar effects
in response functions and asymmetries of the inclusive $\He(e,e')$ reaction around the quasi-elastic
peak. This result is consistent with Ref.~\cite{YUAN201190},
based on the $\Delta$ isobar inclusion in the impulse approximation, as well as 
with a generally observed rather low sensitivity of those observables
to the dynamics beyond the standard forces and currents \cite{golak:05a}.
An example for a sizable and beneficial $\Delta$ isobar effect in the $\He(e,e')$ reaction is
the transverse response function at large three-momentum transfer ${Q} \sim 900 $ MeV
but  low energy transfer $Q_0$ close to the disintegration threshold \cite{deltuva:04b}.
Under those conditions
the cross section and response functions are very small, rendering the measurement highly
challenging. Furthermore, a large value of the three-momentum transfer suggests that
relativistic corrections to the current are not negligible \cite{PhysRevC.82.054003}.
Thus, it would be desirable
to explore the $\Delta$ isobar effect in the region of lower $Q$ values. 
Given the recent and ongoing efforts to study the spin structure of  $\He$,
and thereby also  of the neutron \cite{mihovilovic:14,sulkosky:nature},
the spin observables in the $\He$ electrodisintegration are of a special interest.

Currently the most advanced and quantitative approach to interactions between
nucleons and with electromagnetic probes is the chiral effective field theory.
Its extensions including the $\Delta$ isobar are in progress as well,
however, the $\Delta$ isobar most often is not yet considered as an active degree
of freedom  on the same footing as nucleons, i.e., the derived forces are purely nucleonic.
Few attempts to overcome this restriction
\cite{STROHMEIER2020121980}
are limited to rather low orders of the chiral expansion.  
Therefore the present study
relies on a meson-theoretical but quantitatively most accurate
model for the two-baryon potential with
the $\Delta$ isobar excitation \cite{deltuva:03c}, called CD Bonn + $\Delta$, and the
associated model for the electromagnetic current  \cite{deltuva:04a,deltuva:04b},
containing contributions due to the exchange of
$\pi$, $\rho$, and $\omega$ mesons. In the three-nucleon system
this leads to effective and mutually consistent three-nucleon forces and currents.
The calculations using the purely nucleonic CD Bonn potential are used as reference
to isolate the $\Delta$ isobar effect.

Section II shortly recalls the calculation scheme,
  Sec. III presents the selected results for the electrodisintegration of $\He$,
and Sec. IV summarizes the work. Natural units $\hbar=c=1$
are used throughout the paper.

\section{Calculation scheme \label{sec:eq}}

As customary, the electromagnetic interaction between the electron and the
$\He$ nucleus is treated in the one-photon exchange approximation. The virtual photon couples
to the nuclear electromagnetic current, whose operators are of one-baryon and meson-exchange two-baryon
nature, both purely nucleonic and with the $\Delta$ isobar excitation. They are schematically represented
in Figs.~\ref{fig:Jnn} - \ref{fig:Jdd}  with explicit expressions given in the appendix of 
Ref.~\cite{deltuva:04a}.
Full initial- and final-state interactions for  nuclear states are included
following the Faddeev theory \cite{faddeev:60a,alt:67a} in a more general case 
with channel coupling.
Instead of calculating separately a large number of final states it is convenient
to solve Faddeev-type equations for  auxiliary states 
\begin{equation} \label{eq:X}
    |X^{\lambda} \rangle = {}  ( 1+P)
    (J \cdot \epsilon^\lambda)  | B \rangle  + P T G_0 |X^{\lambda} \rangle,
\end{equation}
where $| B \rangle$ denotes the initial $\He$ bound state,
$J$ is the four-component  electromagnetic current operator,
$\epsilon^\lambda$ is the four-dimensional vector of the virtual photon polarization denoted by $\lambda$, 
$G_0$ is the free resolvent, $T$ is the coupled-channel two-baryon transition
matrix, and $P$ is the sum of two cyclic permutation operators;
see Refs.~\cite{deltuva:04a,deltuva:04b} for more details. The auxiliary states include the action of the
electromagnetic current on  $\He$ bound state and all final-state interactions in the three-nucleon continuum.
Consequently, their projection to nucleon-deuteron or three-nucleon channel states (that are simply free waves)
yields reaction amplitudes and
thereby all observables of two- and three-cluster electrodisintegration
of $\He$ as well as  response functions \cite{deltuva:04a,deltuva:04b}.

Equations (\ref{eq:X}) are solved in the momentum-space partial-wave representation,
including both total isospin $\frac12$ and $\frac32$ components as well as
sufficiently high angular momenta such that the results are well converged;
for example, the total angular momentum up to $\frac{35}{2}$ is taken into account.
The states $|X^{\lambda} \rangle$ have components in all basis states, that is, in the
considered momentum-space partial-wave representation they
depend on two continuous variables,
the Jacobi momenta for the relative motion of three particles, and 
have $NNN$ and $NN\Delta$ components with a large number of orbital angular momentum, spin, and isospin channels,
compatible with the given total angular momentum and parity.
An important conceptual improvement as compared to Ref.~\cite{deltuva:04b}
is the inclusion of the Coulomb interaction between charged baryons using
the method of screening and renormalization \cite{deltuva:05d},
though the results for inclusive observables are barely affected,
except at very low excitation energies near the threshold.

\begin{figure}[t]
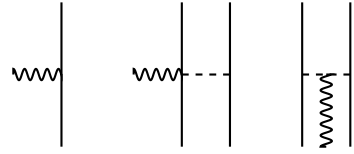

  \begin{center}
     \psset{unit=0.8cm}
\pspicture(6.0,3.0)
\def\nucleon{\psline(0,0)(0,2.4)}
\def\NNNN{\psline(0,0)(0,2.4)\psline(0.8,0)(0.8,2.4) 
\psline[linestyle=dashed,dash=3pt 3pt](0,1.2)(0.8,1.2)}
\def\photon{\pscoil[coilwidth=0.15cm,coilaspect=0,coilarm=0.0cm](0,0)(0.8,0)}
\rput(1,0){\nucleon}
\multips(3,0)(2,0){2}{\NNNN}
\multips(0.2,1.2)(2,0){2}{\photon}
\pscoil[coilwidth=0.15cm,coilaspect=0,coilarm=0.0cm](5.4,0)(5.4,1.2)
\endpspicture
\end{center}
\caption{ \label{fig:Jnn} 
 Purely nucleonic one- and two-baryon electromagnetic currents. 
 Thin solid line denotes the nucleon, the wavy line denotes the photon,
  the dashed line denotes the  instantaneous meson ($\pi,\rho,\omega$) exchange.
  }
\end{figure}

\begin{figure}[!]
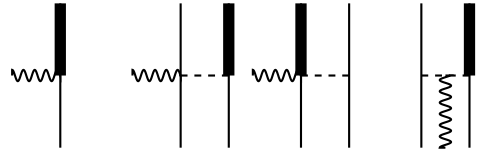

  \begin{center}
    \psset{unit=0.8cm}
\pspicture(8.0,3.0)
\def\nucleon{\psline(0,0)(0,2.4)}
\def\NNNN{\psline(0,0)(0,2.4)\psline(0.8,0)(0.8,2.4) 
\psline[linestyle=dashed,dash=3pt 3pt](0,1.2)(0.8,1.2)}
\def\photon{\pscoil[coilwidth=0.15cm,coilaspect=0,coilarm=0.0cm](0,0)(0.8,0)}
\def\deltaiso{\psline[linewidth=0.15cm](0,0)(0,1.2)}
\rput(1,0){\nucleon}
\multips(3,0)(2,0){3}{\NNNN}
\multips(1.0,1.2)(4,0){2}{\deltaiso}
\multips(3.8,1.2)(4,0){2}{\deltaiso}
\multips(0.2,1.2)(2,0){3}{\photon}
\pscoil[coilwidth=0.15cm,coilaspect=0,coilarm=0.0cm](7.4,0)(7.4,1.2)
\endpspicture
\end{center}
\caption{\label{fig:Jnd}
  One- and two-baryon electromagnetic currents with the excitation of the $\Delta$ isobar,
  that is denoted by a thick line. Hermitean-adjoint contributions are taken into account as well.
}
\end{figure}

\begin{figure}[!]
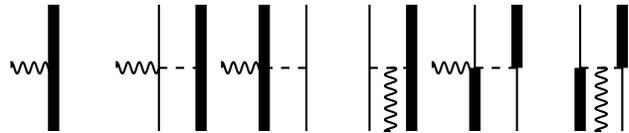

  \begin{center}
    \psset{unit=0.7cm}
\pspicture(12.0,3.0)
\def\nucleon{\psline(0,0)(0,2.4)}
\def\NNNN{\psline(0,0)(0,2.4)\psline(0.8,0)(0.8,2.4) 
\psline[linestyle=dashed,dash=3pt 3pt](0,1.2)(0.8,1.2)}
\def\photon{\pscoil[coilwidth=0.15cm,coilaspect=0,coilarm=0.0cm](0,0)(0.8,0)}
\def\deltaiso{\psline[linewidth=0.15cm](0,0)(0,1.2)}
\def\Deltaiso{\psline[linewidth=0.15cm](0,0)(0,2.4)}
\multips(3,0)(2,0){5}{\NNNN}
\multips(1.0,0.0)(4,0){2}{\Deltaiso}
\multips(3.8,0.0)(4,0){2}{\Deltaiso}
\multips(0.2,1.2)(2,0){3}{\photon}
\pscoil[coilwidth=0.15cm,coilaspect=0,coilarm=0.0cm](7.4,0)(7.4,1.2)
\multips(9.0,0.0)(2,0){2}{\deltaiso}
\multips(9.8,1.2)(2,0){2}{\deltaiso}
\multips(8.2,1.2)(2,0){1}{\photon}
\pscoil[coilwidth=0.15cm,coilaspect=0,coilarm=0.0cm](11.4,0)(11.4,1.2)
\endpspicture
\end{center}
\caption{\label{fig:Jdd}
  One- and two-baryon electromagnetic currents connecting states with a $\Delta$ isobar.
  Due to their negligibly small contributions heavier mesons ($\rho,\omega$) are not included.}
\end{figure} 

\section{Results \label{sec:res}}

Response functions and asymmetries for the  $\He(e,e')$ process with the dynamic
$\Delta$ isobar excitation have been calculated in Refs.~\cite{deltuva:04b,deltuva:05d}
for a limited number of kinematic situations; no significant $\Delta$ isobar effects
have been established except for the near-threshold
transverse response function at high three-momentum transfer.
The results of Refs.~\cite{deltuva:04b,deltuva:05d} remain valid and are not repeated here.
One of the shortcomings in those calculations is the displacement of the quasi-elastic
peak as compared to the experimental data. It is observed also in other works
\cite{golak:05a,PhysRevC.83.057001}
that use nonrelativistic kinematics and dynamics for the three-nucleon system,
since the quasi-elastic scattering conditions calculated with relativistic
and nonrelativistic kinematics deviate from each other with increasing momentum transfer.
Several approaches have been proposed to correct this shortcoming,
such as the use of the active-nucleon Breit frame and the two-fragment model
\cite{PhysRevC.83.057001}.
The present work proposes a simple prescription based on relativistic
\begin{equation}
  \psi_r(Q_0,Q) = (\lambda-\tau) \left \{ \epsilon_F
  \left[ (1+\lambda)\tau + \kappa \sqrt{\tau(1+\tau)} \right] \right\}^{-1/2}
\end{equation}
and nonrelativistic
\begin{equation}
  \psi_{nr}(Q_0,Q) =  \frac{m}{Qk_F} \left ( Q_0 - \epsilon_B - \frac{Q^2}{2m} \right)
\end{equation}
scaling variables taken from  Refs.~\cite{PhysRevC.38.1801,PhysRevC.105.014002}.
Here $\lambda = (Q_0-\epsilon_B)/2m$, $\kappa = Q/2m$, $\tau=(Q^2-Q_0^2)/4m^2$,
$\epsilon_F = [1+(k_F/m)^2]^{1/2}-1$, and  $m$ is the average nucleon mass. 
Parameters $k_F$ and $\epsilon_B$ have the meaning of the Fermi momentum and $\He$
binding energy, their values  are chosen to be $k_F = 180$ MeV and $\epsilon_B= 7.72$ MeV,
though the final results are not sensitive to small variations of $k_F$ and $\epsilon_B$.
The  key point is that experimental and theoretical response functions peak very close to
$ \psi_r(Q_0,Q) = 0$ and $ \psi_{nr}(Q_0,Q) = 0$, respectively, though $(Q_0,Q)$ values in both descriptions
are not identical. Thus, it makes sense to compare the experimental data as a function
of $ \psi_r$ with the nonrelativistic theoretical predictions as functions of $ \psi_{nr}$.
In other words, the experimental data at the given $(Q_0,Q)$ value should be compared with
theoretical predictions taken at slightly modified values $(Q'_0,Q')$  satisfying
$ \psi_r(Q_0,Q) =  \psi_{nr}(Q'_0,Q')$. Obviously, there is some arbitrariness in
choosing two variables $(Q'_0,Q')$ constrained by a single relation, an additional
condition is needed. The present work uses 
$Q' = Q$ though other choices are possible as well.
I emphasize that this is not a rigorous treatment but a prescription that is physicswise 
meaningful in the vicinity of the quasi-elastic peak.
It may be not appropriate near the disintegration
threshold where the excitation energy is the relevant variable. 

\begin{figure}[!]
\begin{center}
\includegraphics[scale=0.6]{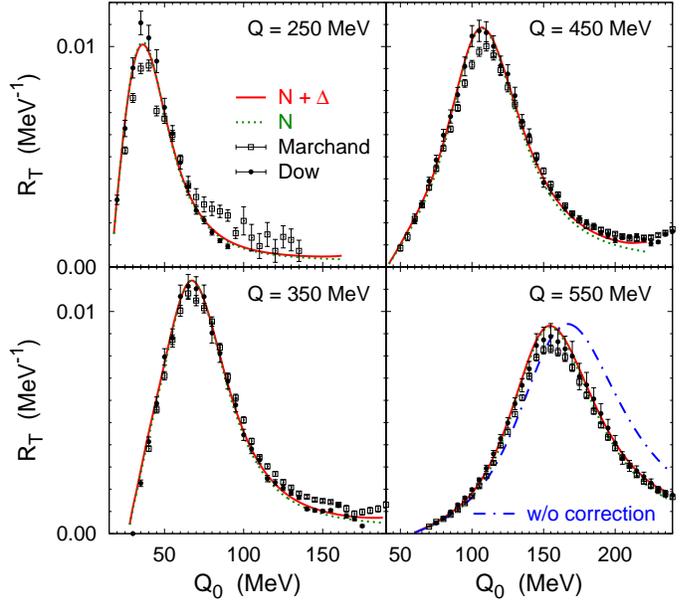}
\end{center}
\caption{\label{fig:rt} (Color online)
Transverse response function $R_T$ as a function of the energy transfer $Q_0$
for momentum transfer values $Q = 250$, 350, 450, and 550 MeV.
Results with and without the $\Delta$ isobar excitation are shown by
solid and dotted curves, respectively.
The dashed-dotted curve at $Q = 550$ MeV displays predictions 
including the $\Delta$ isobar but without the correction for relativistic
 kinematics.
The experimental data are from Refs.~\cite{marchand:85a} $(\Box)$
and \cite{dow:88a} $(\bullet)$. }
\end{figure}

An example applying this prescription is presented in Fig.~\ref{fig:rt} for the
transverse response function $R_T$; it is chosen since it
 is more affected by the $\Delta$ isobar than the longitudinal one.
$R_T$ is shown for  the momentum transfer ranging from  $Q = 250$ to 550 MeV.
 At the highest $Q$ value the results without the above-described correction for the relativistic
 kinematics are included for the comparison. The effect is sizable, the quasi-elastic peak
 is displaced by about 13 MeV in $Q_0$, and the account for the data beyond the peak fails.
 In contrast, with the proposed correction the description of the experimental data is good
 in the whole considered regime,
 especially given the fact that the two sets of data by Marchand et al. \cite{marchand:85a} and
 Dow et al. \cite{dow:88a} are in variance as well; theoretical predictions  favor
 the latter set. The $\Delta$ isobar effect is insignificant around the
 quasi-elastic peak but is more pronounced at larger values of the energy transfer
 beyond the peak, most evident at $Q = 450$ MeV; it is clearly supported by the data.
 This is consistent with the results
 for hadronic scattering \cite{deltuva:03c} where the $\Delta$ isobar effect
 is also enhanced at higher energies.
 The calculations are not pushed to even higher energies for the reason that the underlying
 potentials are not fitted to the two-nucleon data at those higher energies.
The present results are therefore limited to the total relative three-nucleon energy 
below 175 MeV, well below the theoretical $\Delta$ isobar production threshold of 293 MeV.

\begin{figure}[!]
\begin{center}
\includegraphics[scale=0.564]{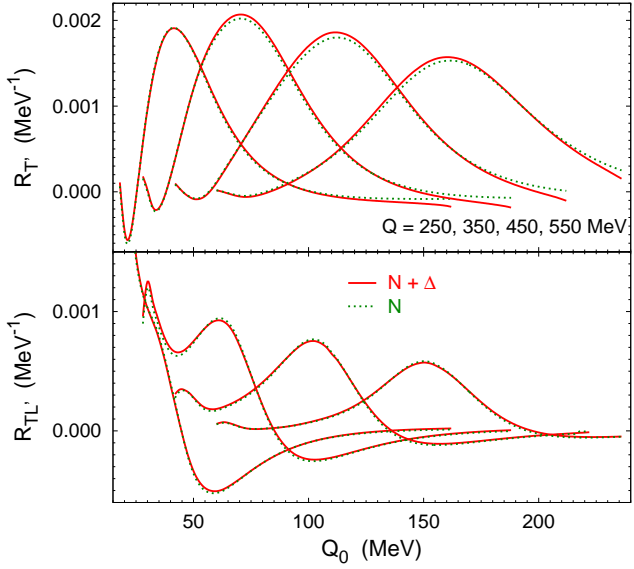}
\end{center}
\caption{\label{fig:rtx} (Color online)
  Transverse and transverse-longitudinal response functions $R_{T'}$ and  $R_{TL'}$ as
  functions of the energy transfer $Q_0$.
Four sets of curves from left to right correspond to
 momentum transfer values $Q = 250$, 350, 450, and 550 MeV.
Curves are as in Fig.~\ref{fig:rt}. 
}
\end{figure}

The spin-dependent transverse and transverse-longitudinal response functions
$R_{T'}$ and  $R_{TL'}$ are studied to a lesser extent. The predictions for the
same values of the momentum transfer are presented in  Fig.~\ref{fig:rtx}. 
To match the convention used in other works, e.g., Ref.~\cite{golak:05a},
the results in the convention
of Ref.~\cite{deltuva:04b} are multiplied by $-1$ and $-\frac12$ for
$R_{T'}$ and  $R_{TL'}$, respectively. The results confirm rapid changes in the shape
of $R_{TL'}$ observed in  Ref.~\cite{golak:05a}.
The $\Delta$ isobar effect appears to be quite insignificant, except for $R_{T'}$
near the quasi-elastic peak and 
at larger values of the energy transfer beyond the quasi-elastic peak,
i.e., the regime with a visible  effect also in $R_T$.

\begin{figure}[!]
\begin{center}
\includegraphics[scale=0.564]{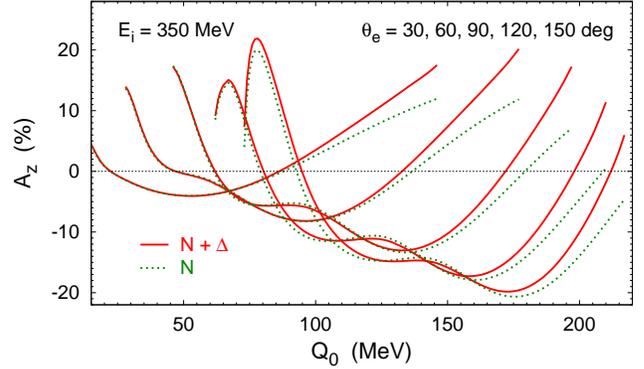}
\end{center}
\caption{\label{fig:az} (Color online)
  The electron polarization  asymmetry for the $\He$ spin oriented parallel to the electron beam
  as a function of the energy transfer $Q_0$.
The electron energy is 350 MeV, five sets of curves from left to right correspond to
electron scattering angles $\theta_e = 30$, 60, 90, 120, and 150 deg.
Curves are as in Fig.~\ref{fig:rt}. }
\end{figure}

Response functions $R_{T'}$ and  $R_{TL'}$ determine the electron polarization asymmetries
for the inclusive $\He(e,e')$ processes \cite{donnelly:86a}.
Calculations for few existing measurements around or below the quasi-elastic peak
\cite{xu:00a,xiong:01a} do not exhibit significant sensitivity to the
three-nucleon force \cite{golak:05a} or $\Delta$ isobar \cite{deltuva:04b}.
The question is under what conditions the $\Delta$ isobar effects shown in
previous figures would show up in the asymmetries that are measurable observables.
From the general asymmetry dependence on the electron kinematics \cite{donnelly:86a}
it is  obvious that the electron
scattering angles $\theta_e$ should take  moderate or large values, otherwise
the asymmetries will be small, just few percent as in Refs.~\cite{xu:00a,xiong:01a}.
Furthermore, to probe the kinematic region around $Q \sim 400$ MeV,  $Q_0 \sim 200$ MeV
with not too small $\theta_e$ 
the electron beam energy $E_i$ should be of few hundred MeV, not in the GeV region.
Few examples of the electron asymmetry predictions under such conditions are presented
in  Fig.~\ref{fig:az}, assuming 350 MeV electron energy,
spin of the $\He$ target oriented in the beam direction,
and five different  electron scattering angles ranging from 30 to 150 degrees.
 The end points of the five sets of curves in Fig.~\ref{fig:az} correspond to
 rather moderate values of $Q = 210$, 303, 377, 431, and 462 MeV,
 the total energies being around or slightly above the pion threshold.
The asymmetry may become as large as 20\% while
at larger energy transfers the inclusion of the $\Delta$ isobar changes it by about
10\% on the absolute scale, e.g., from 7\% to 17\% at $\theta_e = 90$ deg and 
from -5\% to 6\% at $\theta_e = 150$ deg. 
The shown observable is dominated by the $R_{T'}$ contribution which together with
$R_T$ is most affected by the $\Delta$ isobar. Their absolute values well beyond
the quasi-elastic peak are small, but on a relative scale the $\Delta$ isobar effect
is important. A more detailed analysis reveals that the effect is dominated by the
single-baryon current with  $\Delta$ isobar excitation, but $R_{T'}$ gets visible
contribution also from two-baryon meson-exchange currents with  $\Delta$ isobar excitation.
Of course, $NN\Delta$ components in the initial bound and final scattering states are necessary
for the inclusion of those currents, they are generated dynamically by the coupled-channel
potential. Note that for the perpendicular orientation of the target spin
with respect to the beam direction 
the effects are less pronounced, due to the reduced relative weight of $R_{T'}$ and 
increased weight of $R_{TL'}$.

For the experimental measurement of the asymmetry and the verification of the effect
the absolute values of the  cross section are important. Obviously they
are considerably lower than in the vicinity of the quasi-elastic peak.
In the region of interest $Q_0  \succeq 130$ MeV in Fig.~\ref{fig:az} the 
differential cross section decreases with increasing $Q_0$ and
ranges roughly from 1500 to 60 pb/(MeV\,sr), with a moderate dependence on the angle.
Obviously this is lower by several orders of magnitude as compared to
the vicinity of the quasi-elastic peak. On the other hand, it is still
higher by  several orders of magnitude than in the near-threshold measurement
\cite{hicks:03a} that reported values as small as 10 fb/(MeV\,sr).

 
\begin{figure}[!]
\begin{center}
\includegraphics[scale=0.564]{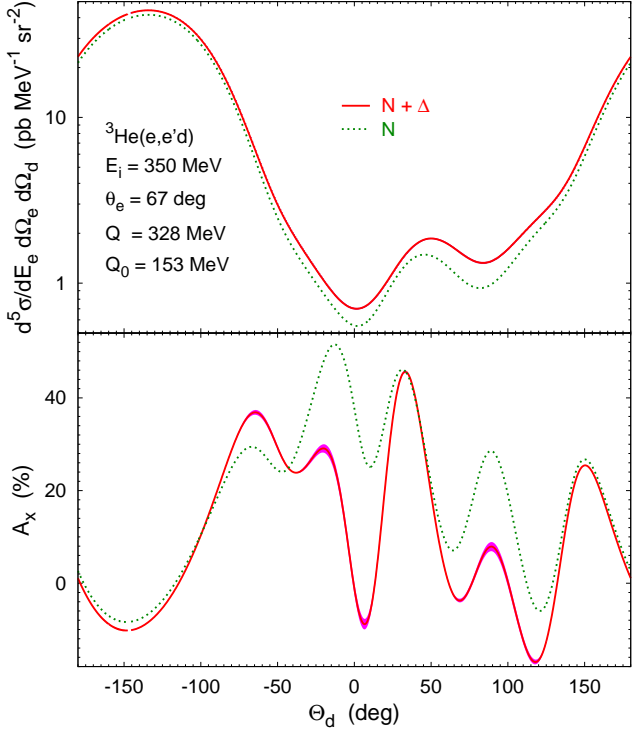}
\end{center}
\caption{\label{fig:axd} (Color online)
  The differential cross section and
  electron polarization  asymmetry for the $\He$ spin in the scattering plane
  oriented perpendicular to the electron beam
  as  functions of the deuteron emission angle $\Theta_d$.
Kinematics conditions are specified in the plot.
Curves are as in Fig.~\ref{fig:rt}. The shaded area reflects the 5\% uncertainty in the
magnetic dipole  form factor $g_{\Delta N}^{M1}(q^2)$.
}
\end{figure}

Given the sizable $\Delta$ isobar effect for the inclusive asymmetry, one may expect even
larger effects  in exclusive kinematics.
An example for the $\He(e,e'd)$ reaction in coplanar kinematics is presented
in  Fig.~\ref{fig:axd}, the total energy is slightly below the pion threshold.
In particular regions of the laboratory deuteron emission angle $\Theta_d$ the
$\Delta$ isobar effect is significant, both for the differential cross section
and the asymmetry. The challenge in the experimental verification again would
be low values of the cross section. On the other hand, the experiment reported
in Ref.~\cite{jans:87a} measured the differential cross section values below
10 pb/(MeV\,sr$^2$). Thus, the measurement of the $\He$ two-cluster electrodisintegration
in the kinematics of Fig.~\ref{fig:axd} might by viable.
The exclusive three-cluster breakup requires the detection of three-particles with even
lower cross section and is even more challenging.

As for the inclusive observables in Figs.~\ref{fig:rt} - \ref{fig:az}, the $\Delta$-isobar effect is largely
dominated by the single-baryon current with the magnetic dipole $\Delta$-isobar excitation, the first
diagram in   Fig.~\ref{fig:Jnd}. The respective form factor $g_{\Delta N}^{M1}(q^2)$ where $q^2=Q^2-Q_0^2$
is taken over from
Ref.~\cite{deltuva:04b} with  $g_{\Delta N}^{M1}(0) = 4.59$ in units of nuclear magneton.
Fits to the experimental data using different dynamical models \cite{PASCALUTSA2007125} yield $g_{\Delta N}^{M1}(0)$
values ranging from 4.52 to 4.61, while 4.71 was employed in Ref.~\cite{ritz:97a}.
Thus, even taking into account uncertainties in the $q^2$-dependence, the overall uncertainty of the presently
used $g_{\Delta N}^{M1}(q^2)$ should not exceed 5\%. The resulting uncertainty for the $\He(e,e'd)$ reaction
observables in Fig.~\ref{fig:axd} is represented by the shaded area around the solid curve and
obviously does not change previous conclusions.

\section{Summary \label{sec:sum}}

The electrodisintegration of the $\He$ nucleus was considered focusing  on the effects
of the $\Delta$ isobar excitation in  electron polarization asymmetries.
The $\Delta$ isobar was treated dynamically in electromagnetic currents and
initial and final hadronic states. Three-particle equations of the
rigorous nonrelativistic Faddeev theory were solved in the momentum-space representation
leading to well-converged results.

The correction for the relativistic kinematics using a simple prescription 
based on the  relativistic and nonrelativistic scaling variables
 places the quasi-elastic peak in the right position and
provides a good description of the experimental data.
The region of the quasi-elastic peak is barely affected by the inclusion of the
$\Delta$ isobar but for larger energy transfer values
both transverse response functions   $R_{T}$ and  $R_{T'}$ show visible effects,
in contrast to longitudinal and longitudinal-transverse ones.
The consequences are predicted sizable  $\Delta$ isobar effects for inclusive electron asymmetries
in particular kinematic regions, typically for few hundred MeV electron beams,
moderate to large electron scattering angles, and energy transfer values
around 150 or 200 MeV. The effects appear even more spectacular for observables
of exclusive processes such as $\He(e,e'd)$.
The experiment performed under similar conditions could provide judgment
for  models of nuclear forces and currents, especially regarding the treatment of the
$\Delta$ isobar.

\vspace{1mm}




\end{document}